\documentclass[aps,prl,twocolumn,showpacs,amsmath,amssymb,amsfonts,nofootinbib,long]{revtex4}
\usepackage{graphicx}

\newcommand\aem{\alpha_{\scriptsize \mbox{em}}}

\newcommand\rot{\nabla\times}

\newcommand\U{U(1)_{\rm PQ}}

\begin{document}
\title{Magnetic Helicity Generation from the Cosmic Axion Field}
\author{L. Campanelli$^{1,2}$}
\email{campanelli@fe.infn.it}
\author{M. Giannotti$^{1,2}$}
\email{giannotti@fe.infn.it}

\affiliation{$^{1}${\it Dipartimento di Fisica, Universit\`a di Ferrara, I-44100 Ferrara, Italy
\\           $^{2}$INFN - Sezione di Ferrara, I-44100 Ferrara, Italy}}

\date{September 28, 2005}


\begin{abstract}
The coupling between a primordial magnetic field and the cosmic
axion field generates a helical component of the magnetic field
around the time in which the axion starts to oscillate. If the
energy density of the seed magnetic field is comparable to the
energy density of the universe at that time, then the resulting
magnetic helicity is about
$|H_B| \simeq ( 10^{-20} \mbox{G})^2 \, \mbox{kpc}$
and remains constant after its generation. As a corollary, we find
that the standard properties of the oscillating axion remain
unchanged even in the presence of very strong magnetic fields.
\end{abstract}


 \maketitle


\section{I. Introduction}

The observations of galaxies and galaxy clusters unambiguously
show that we live in a magnetized universe. Large-scale magnetic
fields with coherence length $\xi \sim \mbox{Mpc}$ and intensities
$B \sim \mu \mbox{G}$ have been observed in all gravitationally
bound large-scale structures. However, whether these magnetic
fields have a primordial origin or were generated during the
period of galaxy formation is still an open question (for a full
discussion see Refs.~\cite{Magn1,Magn2,Magn3,Dol01}).
Understanding this point is a major goal of modern astrophysics
and cosmology, and for this purpose it is extremely important to
consider all the phenomena which leave a different {\it signature}
on magnetic fields of different origin.

As we are going to show below, one of these phenomena is the
interaction of the magnetic field with a cosmic axion field. We
will show in particular that the primordial axion oscillations
generate a helical component of the magnetic field. Of course,
this effect regards only magnetic fields of primordial origin,
since it takes place in the very short interval of time when
axions start to oscillate. This happens when the temperature of
the universe is about $1 \, \mbox{GeV}$, much before the galaxy
formation. In other words, if we believe in the existence of
axions then any {\it primordial} magnetic field observed today
must be (at least partially) {\it helical}.

The magnetic helicity is a very peculiar quantity associated with
a magnetic field \cite{Bis93}, and speculations about the
generation of primordial helical magnetic fields exist in the
literature
\cite{Tur88,Gar92,Fie00,Cor97,Gio98,For00,Vac01,Sem05,Lai05}. In
general, given an electromagnetic field $A^{\mu} = (A^0,{\textbf
A})$ we define the magnetic helicity as
\begin{equation}
\label{Eq1} H_B(t) = \int_V \! d^3 x \, {\textbf A} \cdot \nabla
\times {\textbf A},
\end{equation}
which can be considered as a Chern-Simons term because of the
relation
$\frac{1}{4} \! \int_{t_1}^{t_2} \! d^4 x F_{\mu \nu}
\widetilde{F}^{\mu \nu} = H_B(t_2) - H_B(t_1)$,
where $F_{\mu \nu} = \partial_{\mu} A_{\nu} - \partial_{\nu}
A_{\mu}$ is the electromagnetic field strength tensor,
$\widetilde{F}^{\mu \nu} = (1/2\sqrt{-g}\,)\, \epsilon^{\mu \nu
\rho \sigma} F_{\rho \sigma}$
its dual, $g = \det||g_{\mu \nu}||$ being the determinant of the
metric tensor and $\epsilon^{\mu \nu \rho \sigma}$ the Levi-Civita
(pseudo-)tensor. The definition of the helicity given above is
valid in the Minkowski as well as in the expanding
Freedman-Robertson-Walker universe, since it is easy to see that
Eq.~(\ref{Eq1}) is invariant under general coordinate
transformations. On the other hand, it is odd under discrete $P$
and $CP$ transformations (it is a pseudo-scalar quantity). This
leads to the well known result that the presence of magnetic
helicity in our universe is a manifestation of a macroscopic $P$
and $CP$ violation.

Of course, the phenomenon we are investigating crucially depends
on the existence of the axion field. This is related to the
so-called Peccei-Quinn (PQ) mechanism \cite{Pec77} for the
explanation of the smallness (or absence) of the $CP-$violating
part of the QCD Lagrangian. After almost 30 years, this is still
the most trusted argument for the solution of the above problem,
known as the strong$-CP$ problem (for a general review see, e.g.,
Ref. \cite{Kim87}). This explains why the existence of the axion
is largely believed, even without any experimental evidence for
it, and justifies the great interest in its phenomenology.

Axions are chargeless, weakly interacting, pseudo-scalar particles
which emerge as (pseudo-)Goldstone modes of the (almost) conserved
PQ symmetry $\U$. Although chargeless, axions interact with
photons by means of the anomalous term $g_{a\gamma}a F\tilde F$,
where $g_{a\gamma}=\aem/(2\pi f_a)$, $\aem$ is the electromagnetic
fine structure constant, and $f_a$ is the scale at which $\U$ is
spontaneously broken, known as the axion or PQ constant. This
scale characterizes all the axion properties on the
phenomenological ground. Even though its value is not predicted by
the PQ mechanism, and thus is model dependent, it is constrained
to the very narrow allowed window $10^{10} \lesssim f_a \lesssim
10^{12} \, \mbox{GeV}$ by astrophysical and cosmological
observations and considerations \cite{Tur90,Dav86}. This is a
remarkable point since it will make all our results essentially
independent on the specific axion model.

It is quite interesting (and even rather surprising) that the
axion field, introduced to solve the strong$-CP$ problem in the
QCD Lagrangian, generates a helical component of the primordial
magnetic field, contributing to the macroscopic $CP-$violation in
the universe. This point can be more easily understood considering
the cosmological evolution of the axion field. Axions are born
when the universe is at the temperature $T \sim f_a$, and $\U$
spontaneously breaks. After the universe has cooled down enough
for the axion mass to be comparable with the Hubble expansion rate
$H$, axions start to oscillate driving the QCD Lagrangian to its
$CP-$conserving minimum. During that time axions constitute a
$\mbox{(quasi-)zero}$ momentum condensate, weakly interacting with
the external magnetic field. The main effect of this external
field is to ``extract'' axions from the condensate. This process
is governed by the simple Eq.~(\ref{Eq26b}), where $n_a$ is the
number of axions per co-moving volume and $\varepsilon$ is a
positive, time dependent function, proportional to the magnetic
energy, and inversely proportional to the conductivity of the
primordial plasma. For vanishing magnetic fields, the number of
axion per co-moving volume remains constant during the time of
coherent oscillations. This is a well known result (see, e.g.,
Ref. \cite{Kolb} and references therein) which leads to the upper
limit on the PQ constant (the so-called cosmological bound
\cite{Pre83}). We will see later that this bond remains
essentially unchanged even in the very strong field limit, owing
to the smallness of $\varepsilon$.

However, the effect of the axion primordial oscillations on the
magnetic field is more interesting. In fact, the evaporation of
axions from the condensate is followed by the productions of
photons and a resulting increment of the magnetic field strength.
Since the probabilities of creating left-handed and right-handed
photons are different ($CP$ is not exact), the final result is an
overproduction of photons of one kind with respect to the other.
In other words, the helicity of the system changes.

More specifically, the axion coherent oscillations play the role
of an $\alpha^2-$dynamo on the evolution of the magnetic field
because of the term $\alpha_{\scriptsize \mbox{dyn}} \rot {\textbf
B}$ [see Eq.~(\ref{Eq8})], where $\alpha_{\scriptsize \mbox{dyn}}$
is proportional to $\varepsilon$. This term is $P$ and $CP$ odd,
and is responsible for the generation of the magnetic helicity. As
we expect from the above discussion, for this case the dynamo is
not efficient ($\alpha_{\scriptsize \mbox{dyn}}$ is small), owing
mainly to the smallness of the axion-photon coupling $g_{a\gamma}$
and to the large value of the primordial conductivity $\sigma$.
This last point reflects the known result that helicity is
conserved when the conductivity is infinitively large.

The interaction between pseudo-scalar and electromagnetic fields
has been the object of various papers in the literature.
\\
In the seminal paper by Turner and Widrow \cite{Tur88}, it was
suggested that, during the inflationary epoch, small fluctuating
magnetic fields could have been amplified due to the coupling with
the axion field. This possibility was exhaustively studied in the
subsequent papers by Garretson et al. \cite{Gar92}, and Field and
Carroll \cite{Fie00}.
\\
Ahonen et al. \cite{AER96} studied the evolution of the cosmic
axion field in the background of an external homogeneous magnetic
field just after the QCD phase transition, while in a recent
paper, Lee et al. \cite{Lee02} investigated the possibility of
generating magnetic fields through the coupling of an evolving
pseudo-scalar field during the period of the Large Scale
Structures formation.

In the next Section we will study more accurately the mechanism of
helicity generation by primordial axion oscillations before the
QCD phase transition.

\section{II. Cosmic Axion Oscillations and Magnetic Helicity}

We start with the following interaction Lagrangian between the
cosmic axion field $\phi$ and the electromagnetic field $A_{\mu}$
in curved space-time
\begin{eqnarray}
\label{Eq2} {\mathcal L} \!\!& = &\!\! \sqrt{-g} \left(
\mbox{$\frac{1}{2}$}
\partial_{\mu} \phi \,
\partial^{\mu} \phi - \mbox{$\frac{1}{2}$} m_a^2 \phi^2
- \mbox{$\frac{1}{4}$} F_{\mu \nu} F^{\mu \nu} \right. \nonumber \\
\!\!& + &\!\! \mbox{$\frac{1}{4}$} g_{a \gamma} \phi F_{\mu \nu}
\widetilde{F}^{\mu \nu} + j^{\mu} A_{\mu} \! \left. \right),
\end{eqnarray}
where the external current $j^{\mu}$ takes into account the
interaction between the cosmic plasma and the primordial magnetic
field (see below).

Physics becomes more clear if we introduce the electric and
magnetic fields ${\textbf E}$ and ${\textbf B}$. In a flat
universe described by a Robertson-Walker metric, $ds^2 = dt^2 -
a^2 d{\textbf x}^2$, where $a(t)$ is the expansion parameter
normalized so that at the present time $t_0$, $a(t_0) =1$, this
operation can be performed in the usual way as
\begin{equation}
\label{Eq4} F_{0i} = -a E_i, \;\;\; F_{ij} = \epsilon_{ijk} a^2
B_k,
\end{equation}
where Latin indices range from $1$ to $3$. We shall work in the
Coulomb gauge $A^0 = \partial_i A^i = 0$, in which Eq.~(\ref{Eq4})
become $a {\textbf E} = -\dot{{\textbf A}}$ and $a^2 {\textbf B} =
\nabla \times {\textbf A}$, where a dot indicates the derivative
with respect to the cosmic time $t$, and the spatial derivatives
are taken with respect to co-moving coordinates.

In terms of the electric field, the external current has the form
$j^{\mu} = (0, \sigma {\textbf E})$, where the conductivity of
primordial plasma is a temperature-dependent function that can be
expressed as $\sigma(T) = \kappa T$. For $\Lambda_{\mbox{\tiny
QCD}} \lesssim T \lesssim m_{W}$ (where $\Lambda_{\mbox{\tiny
QCD}} \simeq 200 \, \mbox{MeV}$ and $m_{W} \simeq 80 \,
\mbox{GeV}$), $\kappa$ is a slowly increasing function of
temperature of order unity \cite{Aho96} ($\kappa \simeq 0.76$ for
$T \simeq \Lambda_{\mbox{\tiny QCD}}$, $\kappa \simeq 6.7$ for $T
\simeq m_{W}$). For very high intensities of the magnetic field,
that is above the critical value $B_c = m_e^2/e \simeq 4.4 \times
10^{13} \mbox{G}$ ($m_e$ and $e$ are the mass and absolute value
of the electric charge of the electron), the expression for the
conductivity needs to be multiplied by $B/B_c$ \cite{Enq95}.

Introducing the ``angle'' $\Theta = \phi/f_a$, the Hubble
parameter $H = \dot{a}/a$, and neglecting any spatial variation of
$\phi$, from Lagrangian (\ref{Eq2}) we get the equations of motion
\begin{eqnarray}
\label{Eq5} && \ddot{\Theta} + 3H \dot{\Theta} + m_a^2 \Theta =
\frac{\aem}{2\pi f_a^2} \, {\textbf E} \cdot {\textbf B},
\\
\label{Eq6} && \frac{\nabla \times {\textbf B}}{a} =  {\textbf j}
+ {\textbf j}_D + {\textbf j}_{\Theta},
\end{eqnarray}
where ${\textbf j} = \sigma {\textbf E}$ is the Ohmic current,
${\textbf j}_D = \dot{{\textbf E}} + 2H {\textbf E}$ is the
displacement current in the expanding universe, and we have
introduced the current
${\textbf j}_\Theta = (\aem/2\pi)\dot{\Theta} {\textbf B}$.

It is useful to observe that in the early universe $\sigma \gg H$.
Taking into account the expression for the Hubble parameter $H
\simeq 1.66 g_{*}^{1/2} T^2/m_{Pl}$, where $g_{*}$ is the total
number of effectively massless degrees of freedom and $m_{Pl}$ is
the Planck mass, we get $H/\sigma \sim 10^{-18} \mbox{GeV}/T$
(here we have assumed $B<B_c$, and we have taken $\kappa \sim 1$
and $g_{*} \sim 10^2$). Note that $H/\sigma$ is even smaller for
$B>B_c$.

Observing that $|{\textbf j}_D| / |{\textbf j} \,| \sim H/\sigma$,
in Eq.~(\ref{Eq6}) we can neglect the displacement current with
respect to the Ohmic current. Taking the curl of Eq.~(\ref{Eq6})
we then get
\begin{equation}
\label{Eq8} \dot{{\textbf B}} = -2H {\textbf B} + \frac{\nabla^2
{\textbf B}}{\sigma a^2} + \frac{\aem}{2\pi} \frac{\dot{\Theta}
\nabla \times {\textbf B}}{\sigma a} \, .
\end{equation}
The first term in the right hand side of the above equation
describes the adiabatic dilution of the magnetic field due to the
expansion of the universe, the second one takes into account the
Ohmic dissipations, while the third one violates $P$ and
$CP-$symmetries, and then is responsible for the generation of a
helical component of the magnetic field. Equation (\ref{Eq8})
describes the well known $\alpha^2-$dynamo effect \cite{Magn2} (in
the expanding universe) where the dynamo coefficient is
$\alpha_{\scriptsize \mbox{dyn}} = (\aem/2\pi)
(\dot{\Theta}/\sigma)$. If the dynamo is ``efficient'' (i.e. if
the the dynamo term in the right hand side of Eq.~(\ref{Eq6})
dominates the other two) then the magnetic field is exponentially
amplified, while if the dynamo is not efficient (and neglecting
any dissipative effect) the magnetic field is frozen into the
plasma. We will see that this is indeed our case.

It is useful to work in Fourier space and define the Fourier
transform ${\textbf B}({\textbf k},t)$ of the magnetic field
${\textbf B}({\textbf x},t)$ according to
${\textbf B}({\textbf k},t) = \int d^3 x \, e^{-i {\textbf k}
\cdot {\textbf x}} \, {\textbf B}({\textbf x},t)$,
where ${\textbf x}$ and ${\textbf k}$ are co-moving coordinates
and wavenumbers, respectively. Introducing the orthonormal
helicity basis $\{{\textbf e}_{+}, {\textbf e}_{-}, {\textbf
e}_{3} \}$, where ${\textbf e}_{\pm} = ({\textbf e}_{1} \pm i
{\textbf e}_{2})/\sqrt{2}$, ${\textbf e}_{3} = {\textbf k}/k$
(with $k = |{\textbf k}|$), and $\{{\textbf e}_{1}, {\textbf
e}_{2}, {\textbf e}_{3} \}$ form a right-handed orthonormal basis,
the magnetic field can be decomposed as ${\textbf B} = B_{+}
{\textbf e}_{+} + B_{-} {\textbf e}_{-}$, where $B_{+}$ and
$B_{-}$ represent the positive and negative helicity components of
${\textbf B}$, respectively. In this basis Eq.~(\ref{Eq8}) becomes
\begin{equation}
\label{Eq9} \dot{B}_{\pm} = - 2H B_{\pm} -\frac{k^2
B_{\pm}}{\sigma a^2} \pm \frac{\aem}{2\pi} \frac{\dot{\Theta} k
B_{\pm}}{\sigma a} \, .
\end{equation}
It is clear from the above equation that the two helicity states
of the magnetic field evolve differently due to the $P$ and
$CP-$violating dynamo term. The solutions of Eq.~(\ref{Eq9}) are
easily found
\begin{equation}
\label{Eq10} B_{\pm}(k,t) = B_{\pm}(k,t_i) \left( \frac{a_i}{a}
\right)^{\!2} \exp (-k^2 \ell_{d}^2 \mp k \ell_{\Theta}),
\end{equation}
where $t_i$ is an arbitrary time, $a_i = a(t_i)$, and we have
defined the co-moving dissipation and dynamo lengths
\begin{equation}
\label{Eq11} \ell_{d}^2(t) = \int_{t_i}^{t} \! \frac{dt}{\sigma a}
\, , \;\;\; \ell_{\Theta}(t) = -\frac{\aem}{2 \pi} \!
\int_{t_i}^{t} \! dt \, \frac{\dot{\Theta}}{\sigma a} \, .
\end{equation}
It is useful to introduce the spectra of the magnetic energy and
the magnetic helicity
\begin{eqnarray}
\label{Eq12} {\mathcal E}_B (k,t) \!\!& = &\!\! \left(
\frac{k}{2\pi} \right)^{\!2} \left( \, |B_{+}|^2 + |B_{-}|^2
\right) \! ,
\\
\label{Eq13} {\mathcal H}_B (k,t) \!\!& = &\!\! a^4
\frac{k}{2\pi^2} \left( \, |B_{+}|^2 - |B_{-}|^2 \right) \! ,
\end{eqnarray}
so that the magnetic energy and helicity are given by
\begin{eqnarray}
\label{Eq14} E_B(t) \!\!& = &\!\! \frac{1}{2} \! \int \! d^3 x \,
{\textbf B}^2 = \int \! dk \, {\mathcal E}_B (k,t) ,
\\
\label{Eq15} H_B(t) \!\!& = &\!\! a^2 \!\! \int \! d^3 x \,
{\textbf A} \cdot {\textbf B} = \int \! dk \, {\mathcal H}_B (k,t)
.
\end{eqnarray}
It is clear from Eqs.~(\ref{Eq12}) and (\ref{Eq13}) that any
magnetic field configuration satisfies the realizability condition
$|{\mathcal H}_B (k,t)| \leq 2 a^4 k^{-1} {\mathcal E}_B (k,t)$.
Now, defining
${\mathcal H}^{\mbox{\scriptsize max}}_B (k,t) = 2 a^4 k^{-1}
{\mathcal E}_B (k,t)$,
and inserting Eq.~(\ref{Eq10}) in Eqs.~(\ref{Eq12}) and
(\ref{Eq13}), we get
\begin{eqnarray}
\label{Eq16} && \!\!\!\!\!\!\!\!\!\!\!\!\!\!\!\! {\mathcal E}_B
(k,t) = {\mathcal E}_B (k,t_i) \! \left( \frac{a_i}{a}
\right)^{\!4} \exp (-2k^2 \ell_{d}^2) \cosh (2k \ell_{\Theta}),
\\
\nonumber \\
\label{Eq17} && \!\!\!\!\!\!\!\!\!\!\!\!\!\!\!\! {\mathcal H}_B
(k,t) = - {\mathcal H}^{\mbox{\scriptsize max}}_B (k,t_i) \exp
(-2k^2 \ell_{d}^2) \sinh (2k \ell_{\Theta}),
\end{eqnarray}
where we have supposed that the initial helicity is null. We shall
assume that the initial magnetic energy spectrum can be
represented by the following simple function
\begin{equation}
\label{Eq18} {\mathcal E}_B (k,t_i) = \lambda_B k^p \exp (-2k^2
\ell_{B}^2),
\end{equation}
where $\lambda_B$ and $\ell_B$ are constants. For $k \ll
1/\ell_{B}$, the magnetic energy spectrum possesses a power law
behavior,~
\footnote{Analyticity of the magnetic field correlator $\langle
B_i({\textbf r}_1) B_j({\textbf r}_2) \rangle$ defined on a
compact support forces the spectral index $p$ to be even and equal
or larger than 4 \cite{Cap03}).}
while for large $k$ the spectrum is suppressed exponentially in
order to have a finite energy. Also, we introduce the so-called
correlation length $\xi_B(t)$ defined by
\begin{equation}
\label{Eq19} \xi_B(t) = a(t) \, \frac{\int \! dk \, k^{-1}
{\mathcal E}_B(k,t)} {\int \! dk \, {\mathcal E}_B(k,t)} \, .
\end{equation}
The constant $\ell_B$ is then related to the initial correlation
length $\xi_B(t_i)$ by
\begin{equation}
\label{Eq20} \ell_B = \frac{\Gamma((1+p)/2)}{\sqrt{2} \,
\Gamma(p/2)} \, \frac{\xi_B(t_i)}{a_i} \, ,
\end{equation}
where $\Gamma(x)$ is the Euler gamma function. Then, $\ell_B$ is
proportional to the co-moving initial correlation length.

We now return to the evolution equation for the $\Theta$-angle.
Assuming that the back-reaction [that is the right-hand side of
Eq.~(\ref{Eq5})] of the electromagnetic field is negligible (this
will be justified {\it a posteriori}), the evolution of the cosmic
axion field follows the standard description \cite{Kolb}. Here, we
just remember the expression for the temperature-dependent axion
mass,
$m_a(T) \simeq 0.1 \, m_a(0) (T/\Lambda_{\mbox{\tiny QCD}}
)^{-3.7}$,
valid for $\pi T/\Lambda_{\mbox{\tiny QCD}} \gg 1$, where the
axion mass at zero temperature is
$m_a(0) \simeq 0.6 \, (10^7 \mbox{GeV}/f_a) \, \mbox{eV}$.
The temperature at which the axion field starts to oscillate is
usually indicated by $T_1$, and it is found by solving $m_a(T_1) =
3H(T_1)$. Its value is given by $T_1/\mbox{GeV} \simeq 0.9
f_{12}^{-0.18} \Lambda_{200}^{0.65}$, where $f_{12} = f_a/(10^{12}
\mbox{GeV})$ and $\Lambda_{200} = \Lambda_{\mbox{\tiny QCD}}/(200
\mbox{MeV})$.
\\
Introducing the normalized time $\tau = t/t_1$, the equation of
motion for the $\Theta$-angle becomes
\begin{equation}
\label{Eq22} \Theta'' + \frac{3}{2\tau} \, \Theta' + \frac{9}{4}
\, \tau^{3.7} \Theta = 0,
\end{equation}
where a prime indicates a derivative with respect to $\tau$. In
Fig.~1, we plot the solution of the above equation, where
$\Theta_i = \Theta(t_i)$ and $\dot{\Theta}(t_i) = 0$.


\begin{figure}[h!]
\begin{center}
\includegraphics[clip,width=0.4\textwidth]{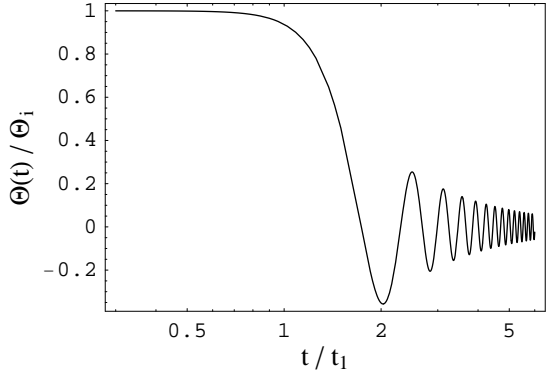}
\caption{$\Theta(t)$ as a function of $t/t_1$.}
\end{center}
\end{figure}


Inserting Eqs.~(\ref{Eq16}) and (\ref{Eq17}) in Eqs.~(\ref{Eq14})
and (\ref{Eq15}) we find, respectively
\begin{eqnarray}
\label{Eq24} \bar{B}^2(t) \!\!& = &\!\! \bar{B}^2_i \left(
\frac{a_i}{a} \right)^{\!4} \left( \frac{\ell_B^2}{\ell_B^2 +
\ell_d^2} \right)^{\!\! (1+p)/2}
\nonumber \\
\!\!& \times &\! _1 F_1 \! \left( \frac{1+p}{2} \, , \frac{1}{2}
\, ; \frac{\ell_\Theta^2 /2}{\ell_B^2 + \ell_d^2} \right) \! ,
\end{eqnarray}
\begin{eqnarray}
\label{Eq25} H_B(t) \!\!& = &\!\! -2 (2\pi)^3 a_i^4 \bar{B}_i^2
\left( \frac{\ell_B^2}{\ell_B^2 + \ell_d^2} \right)^{\!\! (1+p)/2}
\nonumber \\
\!\!& \times &\! _1 F_1 \! \left( \frac{1+p}{2} \, , \frac{3}{2}
\, ; \frac{\ell_\Theta^2 /2}{\ell_B^2 + \ell_d^2} \right)
\ell_\Theta ,
\end{eqnarray}
where $\bar{B}_i = \bar{B}(t_i)$, $_1 F_1 (a,b;x)$ is the Kummer
confluent hypergeometric function \cite{Grad}, and we have
introduced the root-mean-square value of the magnetic field
\begin{equation}
\label{Eq26} \bar{B}^2 = \int \! \frac{d^3k}{(2\pi)^3}
\frac{d^3k'}{(2\pi)^3} \: {\textbf B}({\textbf k}) \cdot {\textbf
B}({\textbf k}') = \frac{2 E_B}{(2\pi)^3} \, .
\end{equation}
From Eq.~(\ref{Eq24}) we see that the dynamo is efficient (and
then there is exponential growth of the magnetic field) if the
dynamo length $\ell_\Theta$ is greater than both the dissipation
length $\ell_d$ and the initial correlation length $\ell_B$.
\\
Magnetic fields whose correlation lengths are smaller than the
dissipation length and for which the dynamo is not efficient are
quickly dissipated. This is easily seen from the evolution
equation for the magnetic field, where the suppression factor due
to dissipation is equal to the third term on the right-hand side
of Eq.~(\ref{Eq24}). Because we are supposing that the magnetic
field interacting with the axion is ultimately the field that we
observe today in the galaxies and galaxy clusters, we assume that
the magnetic field is in the inertial regime, $\ell_d \ll \ell_B$,
during all its evolution.

Since, in general, the conductivity depends on the value of
$\bar{B}$, to know the behavior of the dynamo length $\ell_\Theta$
we have to solve Eq.~(\ref{Eq24}) which is a integral equation for
$\bar{B}$. To this end, we re-write $\ell_\Theta$ as
\begin{equation}
\label{Eq23} \ell_\Theta(\tau) = \frac{\aem}{2 \pi}
\frac{\Theta_i}{\sigma_1 a_1} \, f(\tau),
\end{equation}
where the subscript ``$1$'' indicates that the quantity is
calculated at $t=t_1$, and we have introduced the function
$f(\tau) = -\int_{\tau_i}^{\tau} \! d\tau \tau^{-1/2}
(\Theta'/\Theta_i) (\sigma_1/\sigma)$,
with $\tau_i = t_i/t_1$. We have solved numerically
Eq.~(\ref{Eq24}) for $f(t)$ taking, for purpose of simplicity,
$\sigma = T(1 + \bar{B}/B_c)$, which smoothly interpolates between
the two limiting cases $\bar{B} \ll B_c$ and $\bar{B} \gg B_c$. In
Fig.~2, we plot $f(t)$ as a function of $t/t_1$, where $\Theta$ is
the solution of Eq.~(\ref{Eq22}), for three different cases of
initial strength of the magnetic field. Here, we have
parameterized the initial magnetic field as $\bar{B}_i = b T_i^2$,
so that we have $\bar{B}_i \simeq 1.5 \times 10^{19} b \,
(T_i/\mbox{GeV})^2 \mbox{G}$. [If we force the magnetic energy
density to be less than the energy density of the universe, given
in the radiation era by $\rho = (\pi^2/30) g_{*} \, T^4$, we get
$b \lesssim 0.1 g_{*}(T_i)^{1/2}$.]


\begin{figure}[h!]
\begin{center}
\includegraphics[clip,width=0.4\textwidth]{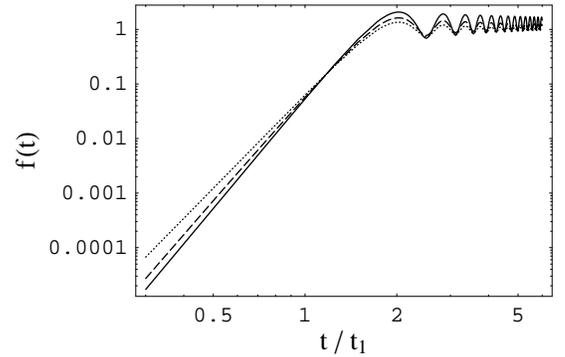}
\caption{$f(t)$ as a function of $t/t_1$ for $T_i = 10 \,
\mbox{GeV}$, $\Theta_i =1$, $\bar{B}_i = bT_i^2$, $\xi_B(T_i) =
10^{-5} d_H(T_i)$, and $p=4$. Solid line: $b = 10^{-1}$,
corresponding to $\bar{B}_i \simeq 10^{20} \mbox{G}$; Dashed line:
$b = 10^{-6}$, corresponding to $\bar{B}_i \simeq 10^{15}
\mbox{G}$; Dotted line: $b = 10^{-10}$, corresponding to
$\bar{B}_i \simeq 10^{11} \mbox{G}$.}
\end{center}
\end{figure}


We checked that, varying the initial conditions for the magnetic
field [i.e. $T_i$, $b$, $\xi_B(T_i)$, and $p\,$], the
characteristic behavior of $f(t)$ in Fig.~2, that is $f(t) \simeq
0$ for $t<t_1$, and $f(t) \simeq 1$ for $t>t_1$, does not change.
This is easily understood if we approximate (see Fig.~1) the
$\Theta$-angle as $\Theta(t)/\Theta_i \simeq 1 - \theta(t-t_1)$,
where $\theta(t)$ is the step-function. In this case, it turns out
that $f$ is a step-function centered at $t=t_1$, $f(t) \simeq
\theta(t-t_1)$.

The dynamo length $\ell_\Theta$ is small compared to the
dissipative length (and then is smaller than $\ell_B$ which we
assumed much greater than $\ell_d$). In fact, for $t > t_1$ and
taking $f \sim 1$, we get $\ell_\Theta / \ell_d \sim 10^{-5}
\Theta_i / \tau^{1/2}$ (where we have taken $\bar{B}<B_c$, and
then the value of $\ell_\Theta / \ell_d$ is even smaller for
magnetic fields for which $\bar{B}>B_c$). Then, from
Eqs.~(\ref{Eq24}) and (\ref{Eq25}), we get that in the inertial
regime the magnetic field evolves as $\bar{B} \propto a^{-2}$
(i.e. the field is frozen into the plasma) while the magnetic
helicity is
\begin{equation}
\label{Eq26'} H_B(t) \simeq -2 (2\pi)^3 a_i^4 \bar{B}_i^2
\ell_\Theta.
\end{equation}
Finally, taking $\ell_\Theta \simeq (\aem/2\pi) (\Theta_i/\sigma_1
a_1)$ for $t>t_1$, we get that the magnetic helicity remains
constant after its generation (say at $t \sim t_1$) and is equal
to
\begin{equation}
\label{Eq27} H_B \simeq -b^2 \Theta_i \, (10^{-17} \mbox{G})^2 \,
\mbox{kpc}
\end{equation}
for $\bar{B}_1<B_c$, corresponding to $b \lesssim 10^{-6}
(\mbox{GeV}/T_1)^2$, and
\begin{equation}
\label{Eq28} H_B \simeq -b \, \Theta_i \! \left(
\frac{\mbox{GeV}}{T_1} \right)^{\!2} (10^{-20} \mbox{G})^2 \,
\mbox{kpc}
\end{equation}
in the case $\bar{B}_1>B_c$, or $b \gtrsim 10^{-6}
(\mbox{GeV}/T_1)^2$.
\\
Since we expect that $\Theta_i$ is of order unity, and because
$T_1 \simeq 1 \, \mbox{GeV}$, we conclude that for very strong
magnetic fields (corresponding to $b \simeq 1$), the generated
magnetic helicity at $t \sim t_1$ is about $|H_B| \simeq (10^{-20}
\mbox{G})^2 \, \mbox{kpc}$.

We can now justify {\it a posteriori} the assumption that the
back-reaction of the electromagnetic field on the evolution of
$\Theta$ is negligible.
\footnote{Below the QCD phase transition, when the axion mass is
constant and equal to its zero temperature value, the fact that a
primordial magnetic field does not interfere with the cosmic
evolution of the axion has been shown in Ref.~\cite{AER96}.}
Integrating Eq.~(\ref{Eq5}) with respect to $d^3x$, observing that
$\dot{H}_B = -2 a^3 \! \int \! d^3 x \, {\textbf E} \cdot {\textbf
B}$,
and taking into account Eqs.~(\ref{Eq11}) and (\ref{Eq26'}), we
obtain
\begin{equation}
\label{Eq26a}  \ddot{\Theta} + (3H  + \varepsilon) \dot{\Theta} +
m_a^2 \Theta = 0,
\end{equation}
where we have defined
$\varepsilon = 2\pi \aem^2 (a_1/a)^4 (\bar{B}_1^2/f_a^2 \sigma)$.
Writing $\bar{B}_1 = b T_1^2$, and assuming $f_a \sim 10^{12} \,
\mbox{GeV}$ and $\bar{B}<B_c$, we get $\varepsilon/H \sim 10^{-10}
b^2 T/\mbox{GeV}$ ($\varepsilon$ is even smaller for
$\bar{B}>B_c$). The smallness of $\varepsilon$ justifies our
previous assumption of neglecting the back-reaction of the
electromagnetic field on the cosmic evolution of the axion.
\\
We finally note that the parameter $\varepsilon$ measures the rate
of the axion condensate evaporation. Indeed, Eq.~(\ref{Eq26a}) can
be written as
\begin{equation}
\label{Eq26b}  \frac{d(\ln n_a)}{dt} = - \varepsilon,
\end{equation}
where $n_a$ is the number of axions per co-moving volume. Thus, we
see that the presence of an external magnetic field opens a new
channel for the dissipation of the axion condensate, though this
dissipation is very small (for other dissipation mechanisms of the
axion condensate see \cite{Lee00} and references therein).

\section{III. Conclusions}

In this paper we have studied in detail the axion-magnetic field
system during the period of primordial axion oscillations ($T\sim
1$ GeV). Here we summarize our main results and give some
perspectives.

The presence of a primordial magnetic field at the time when the
cosmological axion oscillations begin amplifies the axion decay
probability and, contemporaneously, catalyzes the photon
production (enlarging the magnetic energy itself). The effect of
this process on the axion field is minimal, since the number of
decaying axions is rather small $\Delta n_a/n_a\sim 10^{-10}$. A
simple consequence is that the cosmological bound on the axion
mass is preserved, even in the presence of a very strong external
magnetic field.

However, the axion oscillations leave a {\it signature} on the
magnetic field itself, in the form of magnetic helicity. This is
due to the different probability for creating photons of different
helicities. Thus the axion oscillations, interacting with a
magnetic field, originate a macroscopic $CP-$odd state. The amount
of helicity produced depends on the initial intensity of the
magnetic field. At late times, the turbulence of the primordial
plasma should be taken into account in considering the evolution
of both magnetic energy and helicity densities. However, it is now
believed that helicity is a quasi-conserved quantity in
magnetohydrodynamic turbulence \cite{Bis93,Tay74}. Hence, the
helicity generated in the primordial universe should survive until
today without changing its major properties, and could be
important for the dynamo mechanism of magnetic fields operating in
galaxies and clusters of galaxies (see, e.g., Ref. \cite{Magn1}).
Moreover, the presence of a helical component may speed up the
growth of the magnetic field correlation length during its
evolution in the primordial universe through the so-called inverse
cascade of magnetic energy \cite{Bra96}, leading today to magnetic
fields on larger scales \cite{Fie00,Son99}.

In principle, a helical magnetic field could leave peculiar
imprints on the Cosmic Microwave Background Radiation (CMBR).
Unfortunately, the maximal helicity producible in our mechanism,
$|H_B| \simeq (10^{-20} \mbox{G})^2 \, \mbox{kpc}$, is much
smaller than that detectible in near future CMBR experiments,
which is of order of $(10^{-9} \mbox{G})^2 \, \mbox{Mpc}$
\cite{Cap04}. However, from the above discussion it emerges a
strong relation between axion and magnetic helicity. We believe
that this subject deserves further investigations.

\vspace*{0.2cm}

\begin{acknowledgments}
We would like to thank A. D. Dolgov for illuminating discussions
and for carefully reading the manuscript. We also thank F. L.
Villante, D. Comelli, A. Drago, and F. Nesti for helpful
discussions.
\end{acknowledgments}



\end{document}